\documentclass[preprint,prl,amsmath,amssymb]{revtex4-1}

\usepackage{graphicx}
\usepackage{mathptmx}

\usepackage{color}
\usepackage{ulem}

\begin{document}

\title{Playing graphene nanodrums: force spectroscopy of graphene on Ru(0001)}

\author{Elena Voloshina$^{1,}$\footnote{Corresponding author. E-mail: Elena.Voloshina@hu-berlin.de} and Yuriy Dedkov,$^{2,}$\footnote{Corresponding author. E-mail: dedkov@ihp-microelectronics.com, Yuriy.Dedkov@icloud.com; Present address: IHP, Im Technologiepark 25, 15236 Frankfurt (Oder), Germany}}

\affiliation{\mbox{$^1$Humboldt-Universit\"at zu Berlin, Institut f\"ur Chemie, 10099 Berlin, Germany}}

\affiliation{\mbox{$^2$SPECS Surface Nano Analysis GmbH, Voltastra\ss e 5, 13355 Berlin, Germany}}

\date{\today}

\begin{abstract} 
Graphene, a thinnest material in the world, can form moir\'e structures on different substrates, including graphite, $h$-BN, or metal surfaces. In such systems the structure of graphene, i.\,e. its corrugation, as well as its electronic and elastic properties are defined by the combination of the system geometry and local interaction strength at the interface. The corrugation in such structures on metals is heavily extracted from diffraction or local probe microscopy experiments and can be obtained only via comparison with theoretical data, which usually simulate the experimental findings. Here we show that graphene corrugation on metals can be measured directly employing atomic force spectroscopy and obtained value coincides with state-of-the-art theoretical results. We also address the elastic reaction of the formed graphene nanodoms on the indentation process by the scanning tip that is important for the modeling and fabrication of graphene-based nanoresonators on the nanoscale.
\end{abstract}

\maketitle

Graphene superlattices~\cite{Ponomarenko:2013hl,Hunt:2013ef,Yankowitz:2012gi,Sutter:2014kc}, and among them graphene-based moir\'e structures, attract increased attention in the recent condensed matter studies because they can be used as a playground for tailoring the transport properties of the graphene-based systems~\cite{Ponomarenko:2013hl,Hunt:2013ef,Yankowitz:2012gi}. Formation of the ordered commensurate moir\'e structures of graphene on different substrates leads to the cloning of Dirac cones in the reciprocal space that in its turn modifies the electronic spectrum of carriers via opening of mini-gaps around the Dirac point, deviates the energy band dispersion from the linear one, changes the effective mass and velocity of the carriers, etc.~\cite{Ponomarenko:2013hl,Hunt:2013ef,Yankowitz:2012gi,Sutter:2014kc,Pletikosic:2009,Rusponi:2010}. All these effects are a manifestation of the additional large-scale modulation potential originating from the moir\'e superlattice, which usually has a size of several nanometers.

The existence of moir\'e structures on the basis of graphene is known for many years and they were successfully imaged by means of scanning tunneling microscope (STM) for graphene on common graphite~\cite{Xhie:1993ub,Rong:1993um} and graphene on Si- or C-faced SiC~\cite{Lauffer:2008cb}. Here graphene moir\'e structures appear naturally during preparation of these systems for experiment due to the weak interaction between single graphene layers. The periodicities as well as corrugations of such graphene moir\'e structures are defined by the misalignment angle between two carbon lattices in the neighboring graphene layers. The recent experiments on the transport properties of the ultraflat graphene in the artificial graphene/h-BN heterostructures reveal the modified mass-less electronic spectrum of graphene. The appearance of the moir\'e-lattice related Dirac points in the electronic structure of graphene is manifested as local dips in the STM spectroscopy data~\cite{Yankowitz:2012gi} ($dI/dV$ signal is proportional to the local density of states) and via observation of the so-called Hofstadter butterfly ``self-similar'' superlattice energy spectrum~\cite{Ponomarenko:2013hl,Hunt:2013ef} for the charge particle moving under simultaneous influence of two periodic potentials, atomic-lattice- and moir\'e-lattice-related, and magnetic field.

One of the moir\'e lattices classes is a graphene layer on the close-packed surfaces of $4d$ and $5d$ transition metals, like Ir(111), Pt(111), or Ru(0001). Such graphene-metal systems are the subject of the long-term surface science studies and they were proposed as substrates for the ordered arrays of metallic clusters, which then can be used in storage technology or in catalysis as the behaviour of a clusters' array could be modelled on the basis of a single element. The electronic or magnetic properties of such cluster arrays strongly depend on the underlying graphene-metal substrate. The crystallographic structure of these graphene-metal systems, and hence their electronic properties, are defined by the lattice mismatch of graphene and metal surface as well as by the strength of the local interaction at the interface, leading to the observation of various moir\'e structures with different corrugations even for the same graphene-metal combinations~\cite{Busse:2011,Stradi:2011be,Voloshina:2012a}. Thus the precise knowledge of the crystallographic structure of the graphene-metal system is crucial for the modelling of its electronic properties. Macroscopic diffraction experiments, like low-energy electron diffraction (LEED), or local probe methods, like STM and atomic force microscopy (AFM), can give information about symmetry of the system and lattice alignment of graphene and metal~\cite{Moritz:2010,Voloshina:2013dq,Hamalainen:2013jj}. The graphene moir\'e lattice corrugation on metal is extracted from the comparison of experimental and theoretical data, e.\,g. from the modelled STM images at different bias voltages~\cite{Stradi:2011be,Voloshina:2013dq,Stradi:2012hw} or from simulated $I(V)$ curves in LEED experiments~\cite{Moritz:2010,Hamalainen:2013jj}. AFM in principle can do this job, but as was shown, the correct topography in this case is influenced by the residual electrostatic forces between tip and sample~\cite{Sadewasser:2003jg}. In such AFM experiments, the compensation of the local contact potential difference variation over the moir\'e lattice is not a trivial task. Also, because graphene is very elastic material, the imaged topography in AFM experiments can be influenced by the indentation effect from the tip, which is placed in the close vicinity to the surface~\cite{Boneschanscher:2014dp,Koch:2013bk}. Therefore, the necessity of the tool, which gives full information about graphene moir\'e, is obvious and our approach is based on the complementary STM and AFM spectroscopy measurements in this case.

Let us consider a graphene layer on Ru(0001) as the most representative case of the strongly corrugated graphene-metal system with the alternating places of the different stacking of graphene on metal substrate. Here high-symmetry places of the moir\'e structure are called with respect to the adsorption places of Ru(0001) which are centres of the carbon rings: ATOP, FCC, and HCP (see inset of Fig.\,1a). Such different local atom stackings of carbon and metal atoms lead to the modulation of the interaction strength at the interface yielding the formation of a strongly corrugated graphene layer on Ru(0001) with the height difference between two limit positions of $127$\,pm as deduced from our DFT calculations. The natural approach to use STM or AFM to measure true corrugation does not work here as the variation of the local density of states leads to the strong dependence of the apparent corrugation on the bias voltage between tip and sample~\cite{Stradi:2011be,Voloshina:2013dq,Stradi:2012hw} or graphene nanodoms can be modified via the local chemical interaction with the apex of the tip~\cite{Boneschanscher:2014dp,Koch:2013bk}.

If scanning metallic tip is placed above one of the high-symmetry positions of the graphene moir\'e structure then one can expect that the interaction energy between tip and graphene will be described by Morse or Lenard-Jones (LJ) potential (see Fig.\,S1 and the respective text in the Supplementary material for the discussion). Both of them consist of the attractive and repulsive parts, which are more effective on long and short distances between objects, respectively. We calculated the interaction energy between our model tip and graphene/Ru(0001) varying the distance between apex of the tip and graphene (for details, see Supplementary material). Here distance $z$ denotes the interval between middle W layer of the model tip and middle Ru layer in the graphene/Ru(0001) slab. Such computational experiments model the real AFM force-spectroscopy measurements. For HCP or FCC places we obtain the expected curves for the tip-sample interaction energy (top panel of Fig.\,1a), which have a shape very close to the one for Morse or LJ potential. The minimum value for the interaction energy is $E=-1.26$\,aJ at $z=1$\,nm.

Surprisingly, the calculated curve for the interaction energy for the ATOP place shows a clear depression around $z=1.2$\,nm on the attractive part (point 4 in Fig.\,1a; distance between W-atom of the tip apex and C-atom underneath is $227$\,pm) followed by the global minima for the energy of $E=-1.02$\,aJ at $z=1.01$\,nm (point 6 in Fig.\,1a; distance between W-atom of the tip apex and C-atom underneath is $211$\,pm). This effect is explained by the drum-like behavior of the ATOP places of graphene moir\'e on Ru(0001). If tip is approaching the graphene bubble at the ATOP position, first it interacts with a graphene layer and this interaction can be described by Morse or LJ potential and can be assigned to the formation of the W-C bond at the short distance between apex of the tip and graphene layer (Fig.\,1b, panels 1-4). However it is possible to press with the tip on a graphene layer further until the strong repulsive interaction between graphene and the underlying Ru prevent further graphene flexure. This complex behaviour leads to the appearance of the global minimum on the curve for the interaction energy between tip and graphene at the ATOP position and leads to the formation of several W-C bonds between tip and graphene (Fig.\,1b, panels 4-6). Movie\,S1 (Fig.\,S2) of the Supplementary material presents a series of snapshots of the system geometry, overlaid with the charge density distribution, obtained during indentation process. Decomposition of the total curve for the ATOP position on two curves describing Morse or LJ potential is not possible as it involves interaction between three objects, a scanning tip, a graphene layer, and the Ru(0001) substrate. Differentiation of the curve for the interaction energy, $F_z=-dE/dz$, gives the force curves, which acts on the scanning tip, for different places of the graphene/Ru(0001) moir\'e structure (bottom panel of Fig.\,1a) and these curves can be directly compared with the results of AFM spectroscopy experiments.

To verify these theoretical predictions the combined STM/AFM experiments were performed. We prepared graphene on Ru(0001) in the usual fashion via decomposition of the ethene gas at $1020$\,K. The representative STM topography of the as prepared graphene/Ru(0001) moir\'e structure is shown in the inset of Fig.\,2a (see also Fig.\,S3 of the Supplementary material), where all high-symmetry positions are clearly resolved and identified. These results are in perfect agreement with our simulated STM images obtained in the framework of the Tersoff-Hamann formalism~\cite{Tersoff:1985} on the basis of DFT calculations (Fig.\,S4 of the Supplementary material). Our combined scanning probe microscopy/spectroscopy experiments were performed with the oscillating conductive tip (resonance frequency $f_0$) allowing to measure tip-graphene interaction forces as a function of distance. Interaction between tip and surface leads to the slight change of the resonance frequency and this frequency shift $\Delta f(z)$ signal is measured, which is later transformed in the component of the force parallel to oscillations, $F_z(z)$, and interaction energy, $E(z)$, between tip and graphene~\cite{Sader:2004kt}. Such measurements can be performed on the $xy$-grid on the sample surface producing the 3D data sets, $\Delta f(x,y,z)$, that allows the careful analysis of the data (see Fig.\,S5 and Movie\,S2 of the Supplementary material). Figure 2a (top panel) shows representative $\Delta f(z)$ curves extracted from such data sets for the middle positions of the respective high-symmetry places. The corresponding forces acting on the tip and the respective tip-graphene interaction energy curves at different places of the moir\'e structure are presented in the middle and the bottom panels, respectively. All sets show a clear depression on the attractive part of the curves for the ATOP positions. We can indicate the extremely good agreement between experimental and theoretical curves for $F_z(z)$ and $E(z)$. Simultaneously with the   signal, we also perform measurements of the so-called dissipation signal, $\Delta U(z)$, which indicates the energy need in order to keep the oscillation amplitude constant (Fig.\,2b; see also Fig.\,S5 and Movie\,S3 of the Supplementary material). Several dissipative mechanisms might be responsible for this effect, but here, the most relevant one is the excitation of the movement of the graphene sheet due to the attractive interaction between graphene and oscillating tip. One can clearly recognise two bumps on the curve for the dissipative channel for the ATOP places, each of them is connected with the respective minima in the curve for the interaction force. The difference in the value for the dissipation energy between ATOP and HCP/FCC places is due to the weaker interaction between graphene and Ru for the ATOP place compared to HCP/FCC, that causes the smaller energy loss for keeping oscillation amplitude constant.

Comparison of the obtained experimental results with the previously discussed DFT data shows very good agreement between theory and experiment (keeping in mind the complexity and size of the studied system consisting of 1140 atoms in the DFT simulations). Analysing the $\Delta f(z)$ signal we can take the difference between the positions of the first minima in the curve for the ATOP place and of the minima for the HCP place as a corrugation of the graphene on Ru(0001) that yields a value of $130$\,pm, which coincides with $127$\,pm obtained in DFT calculations. 

Analysis of the force curves shows that subtraction of the long-range tail (originating from the macro-tip) from the curve for the ATOP place gives for the first minima (in the first approximation) the reaction of the graphene nanobubble on the pressure from the tip side. Linear fit of the initial retraction part of this curve gives a stiffness of the graphene layer of $6.88$\,N/m and finally the resonant frequency of approximately $0.74$\,THz (see Supplementary material for details). This value can be compared with the similar values of $10.6$\,N/m and $0.92$\,THz, for graphene stiffness and resonance frequency of graphene-bubbles, respectively, obtained from the theoretical data. Of course, the validity of the elastic theory for the continuous media might be questionable on the nanoscale and further theoretical analysis is necessary. However, even this simplified approach gives very good agreement between experimental and theoretical data.

The reproducibility of the observed effect is demonstrated in Fig.\,2c (see also Figs.\,S5 and S6, Movie\,S2 and Movie\,S3 of the Supplementary material), where several maps for the $\Delta f(z)$ and $\Delta U(z)$ channels measured in the constant height AFM mode for the graphene moir\'e lattice on Ru(0001) are shown, i.\,e. for the array of graphene nanoresonators. The cuts presented in Fig.\,S5 were extracted from the 3D sets of data, which were collected on the $xy$-grid as discussed earlier. One can clearly see that all graphene nanobubbles in ATOP places show the same nanodrum behaviour with the same elastic properties. The similar effect was also observed in the pure constant height data for the $\Delta f(z)$ and $\Delta U(z)$ channels as shown in Fig.\,2c and Fig.\,S6 of the Supplementary material. As was demonstrated~\cite{Pan:2009vv} such assembly of graphene nanoresonators can be fabricated on the mm-size substrates, where all these nm-sized nanoresonators have the same characteristics and performance, that opens the high perspectives for application of such nanoresonators-arrays in THz communication and biosensing~\cite{Tassin:2013jw}.

\section*{Methods}

\subsection{DFT calculations.}

The DFT calculations were carried out using the projector augmented wave (PAW) method~\cite{Blochl:1994}, a plane wave basis set and the generalized gradient approximation as parameterized by Perdew \textit{et al.}~\cite{Perdew:1996}, as implemented in the VASP program~\cite{Kresse:1994}. The plane wave kinetic energy cutoff was set to $400$\,eV. The long-range van der Waals interactions were accounted for by means of the DFT-D2 approach~\cite{Grimme:2006}. The supercell used to model the graphene-metal interface has a ($12\times12$) lateral periodicity with respect to Ru(0001). It is constructed from a slab of 3 layers of Ru atoms with a ($13\times13$) graphene layer adsorbed from both sides. The W-tip is modelled with a 32-atom (i.\,e. $1-4-9-4-9-4-1$) cluster. Thus our structural model (see Fig.\,S1 of the Supplementary material) is completely symmetric with respect to the middle-layer of the slab and middle-layer of the W-cluster. During the structure relaxation, the positions of the carbon atoms ($x,y,z$-coordinates) as well as these of the apex W atoms ($z$-coordinate) are allowed to relax. The surface Brillouin zone is sampled with a single $k$-point at the $\Gamma$-point. The STM images are calculated using the Tersoff-Hamann formalism for the fully relaxed graphene/Ru(0001) slab with 5 Ru layers and ($3\times3\times1$) sampling in the Brillouine zone.

\subsection{Sample preparation.}

Prior to every set of experiment, the Ru(0001) was prepared via cycles of Ar-ion-sputtering and annealing procedure. Graphene/Ru(0001) system was prepared in ultrahigh vacuum system via cracking of ethylene: $T=1020$\,K, $p=2\times10^{-7}$\,mbar, $t=30$\,min. This procedure leads to the single-domain graphene layer on Ru(0001) of very high quality, which was verified by means of STM and AFM (Fig.\,S3 of Supplementary material).

\subsection{STM and AFM experiments.}

The STM and AFM measurements were performed in constant current, constant frequency shift, or constant height modes, respectively. In the first two cases the topography of sample, $z(x, y)$, is studied with the corresponding signal, tunnelling current ($I_T$) or frequency shift ($\Delta f$), used as an input for the feedback loop. In the last case the feedback was completely switched off and $I_T$, $\Delta f$, $\Delta U$ signals were collected in the atom-tracking mode allowing for the thermal drift compensation for the tip $z$-position. $z$-spectroscopy data were performed on the grid of ($96\times96$) or ($128\times128$) pixels and topography of the sample was recorded simultaneously allowing to obtain the ``$z=0$'' reference point used in the data-treatment as well as for the careful tracing of the drift in the $xy$-plane. All STM/AFM images were collected at room temperature with SPM Aarhus 150 equipped with KolibriSensor$\texttrademark$ from SPECS GmbH with Nanonis Control system. In these measurements the sharp W-tip was used which was cleaned in situ via Ar$^+$-sputtering. In presented STM images the tunnelling bias voltage, $U_T$, is applied to the sample and the tunnelling current, $I_T$, is collected through the tip, which is virtually grounded. During the AFM measurements the sensor was oscillating with the resonance frequency of $f_0=998666$\,Hz and the quality factor of $Q = 22500$. The oscillation amplitude was set to $A = 200$\,pm or $A = 300$\,pm. Base pressure during all experiments was below $8\times10^{-11}$\,mbar.

\section*{Acknowledgements}

The High Performance Computing Network of Northern Germany (HLRN-III) is acknowledged for computer time. Financial support from the German Research Foundation (DFG) through the grant VO1711/3-1 within the Priority Programme 1459 ``Graphene'' is appreciated.

\section*{Author contributions}

E.V. performed DFT calculations. Y.D. carried out STM/AFM experiments. Both authors contribute equally in the analysis of data and writing of the manuscript.

\section*{Additional information}

\textbf{Competing financial interests:} The authors declare no competing financial interests.

\clearpage
\begin{figure}[t]
\includegraphics[width=0.9\textwidth]{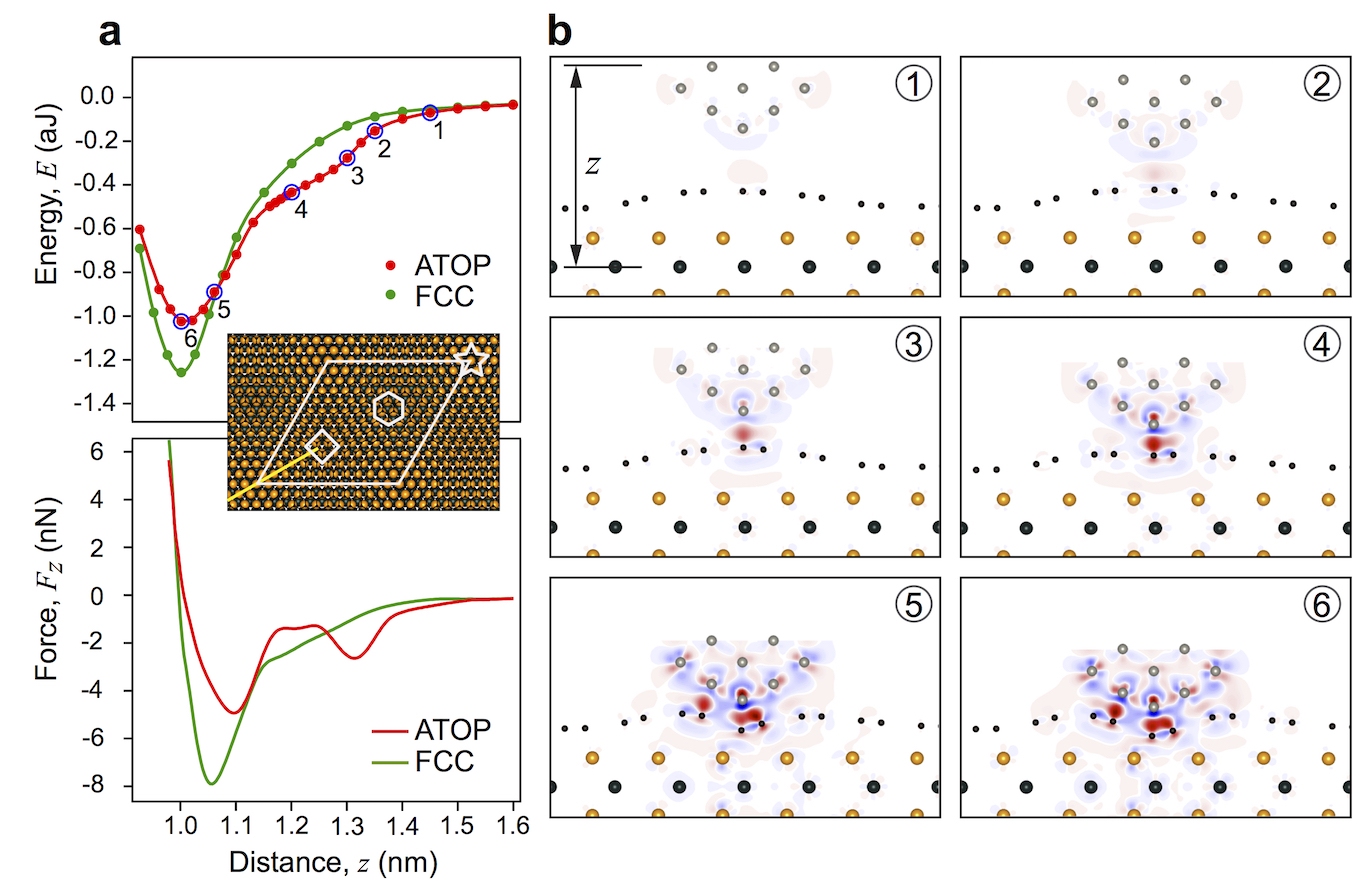}\\
  \caption{Theoretical modelling of the tip-graphene interaction in AFM.  (\textbf{a}) Interaction energy ($E$) calculated at two different places of graphene/Ru(0001) (top panel) and the respective tip-sample interaction force ($F_z$) (bottom panel) as functions of the tip-sample distance ($z$). (\textbf{b}) Series of snapshots of the atomic configurations corresponding to the interaction of W-tip with graphene/Ru(0001) at the ATOP place taken for six different distances indicated with numbers 1-6 at top panel of (\textbf{a}). The cut is made perpendicular to the sample surface through the yellow line as shown in the inset of (\textbf{a}). Each structure is overlaid with the calculated difference electron density, $\Delta\rho=\rho_{\mathrm{tip+sample}}(r)-[\rho_{\mathrm{tip}}(r)+\rho_{\mathrm{sample}}(r)]$. Red (blue) colour indicates accumulation (depletion) of the electron density. The scale is identical for each subplot and ranges from $-0.03 e/\mathrm{\AA}^3$ to $+0.03 e/\mathrm{\AA}^3$.}
\end{figure}

\clearpage
\begin{figure}[t]
\includegraphics[width=0.9\textwidth]{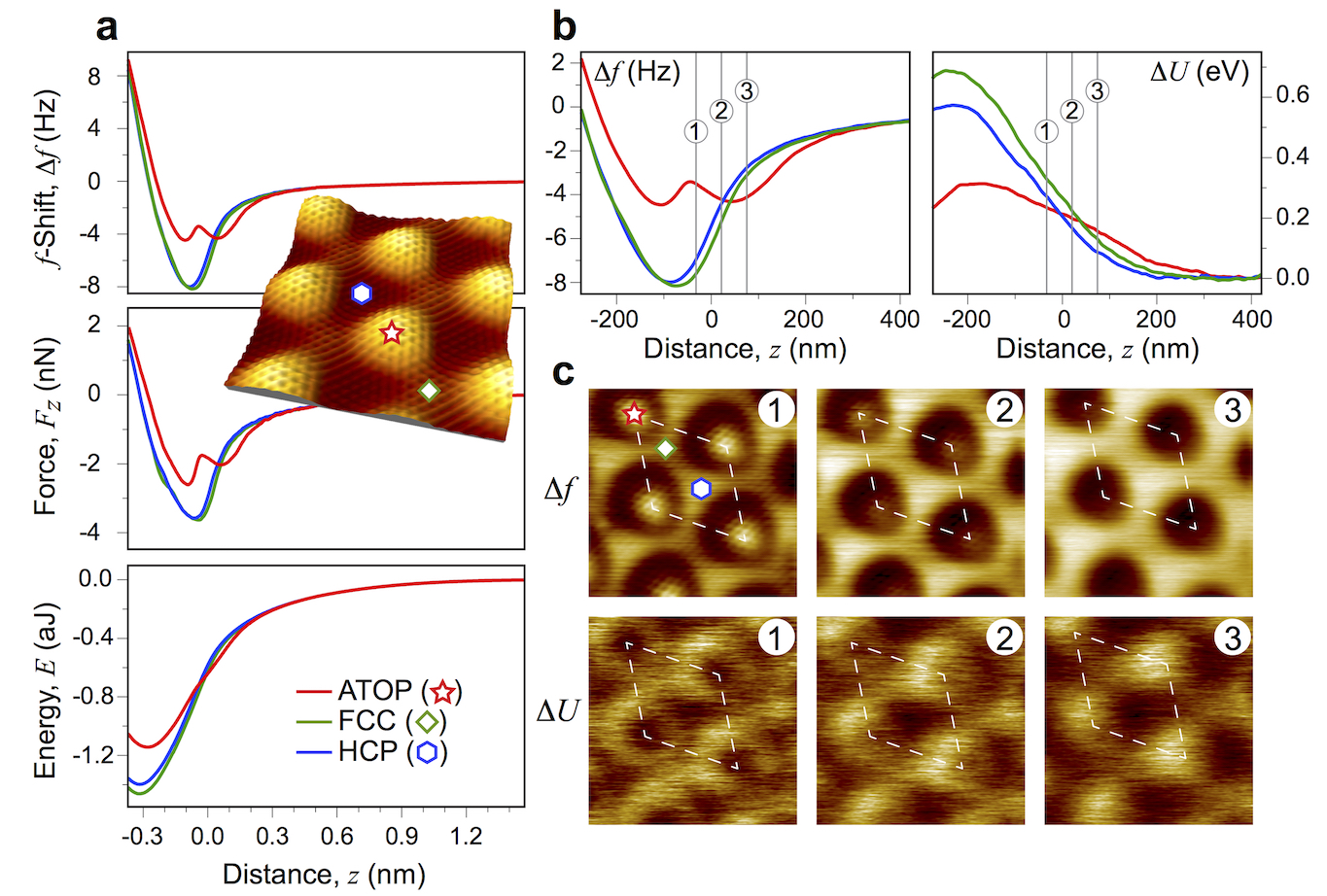}\\
  \caption{Force-distance spectroscopy and constant-height SPM of graphene/Ru(0001). (\textbf{a}) Frequency shift as a function of the tip-sample distance ($\Delta f$) measured at different places of graphene/Ru(0001) (top panel) and the respective tip-sample interaction force ($F_z$) and energy ($E$) are shown in the middle and the bottom panels. Inset shows the 3D view of the STM image of graphene/Ru(0001) with high-symmetry positions of the moir\'e structure marked by the corresponding symbols. Imaging parameters: $6.4\times6.4\mathrm{nm}^2$, $U_T = +0.05$\,V, $I_T = 5$\,nA. (\textbf{b}) Frequency shift ($\Delta f$) and dissipation ($\Delta U$) signals zoomed for the tip-sample distances corresponding to the short-range interactions. (\textbf{c}) Series of constant-height SPM images ($\Delta f$ and $\Delta U$ signals) measured at the $z$-positions marked in (\textbf{b}). Minimum (dark) and maximum (bright) limits for the respective measurement channel can be extracted from the panel (\textbf{b}). Imaging parameters: $6.4\times6.4\mathrm{nm}^2$, $U_T = +0.05$\,V, $A = 200$\,pm.}
\end{figure}

\clearpage

\noindent
Supplementary material for manuscript:\\
\textbf{Playing graphene nanodrums: force spectroscopy of graphene on Ru(0001)}\\
(movie available upon request)\\
\newline
Elena Voloshina$^1$ and Yuriy Dedkov$^{2}$\\
\newline
$^1$Humboldt-Universit\"at zu Berlin, Institut f\"ur Chemie, 10099 Berlin, Germany\\
$^2$SPECS Surface Nano Analysis GmbH, Voltastra\ss e 5, 13355 Berlin, Germany\\
\\
\noindent
\textbf{Interaction energy.} When studying the interaction between W-tip and the graphene/Ru(0001) sample, the interaction energy can be written as 
 \begin{equation}
\Delta E_{\mathrm{int}}(z)=E_{\mathrm{tip+sample}}(z)-[E_{\mathrm{sample}}(z)+E_{\mathrm{tip}}(z)],
\end{equation}
where $E_{\mathrm{sample}}(z)$ and $E_{\mathrm{tip}}(z)$ are the energies of the isolated tip and sample, and $E_{\mathrm{tip+sample}}(z)$ is the energy of their interacting assembly.

There are two different ways to calculate $E_{\mathrm{sample}}(z)$ and $E_{\mathrm{tip}}(z)$. In the first one the structures of isolated tip and sample are fully relaxed. This way, $E_{\mathrm{sample}}(z)$ and $E_{\mathrm{tip}}(z)$ represent constant functions of distance between the tip and the sample and the interaction energy curve have the same qualitative performance as the total energy of the interacting assembly versus the tip-sample distance. Alternatively, one can freeze the structures of the isolated tip and sample as obtained after the relaxation for each fixed tip-sample distance. This way, the energies of the tip and the sample will represent functions of distance and the performance of the interaction energy curve will depend on the strength of the tip-sample interaction: The stronger interaction (the more significant modification the tip and the sample undergo) yields stronger modification of the qualitative performance of the interaction energy curve relative the performance of the total energy of the interacting assembly versus the tip-sample distance.

In this work we employed the second approach, i.\,e. after initial relaxation of the tip-sample geometry, the structures of the isolated tip and sample for each fixed tip-sample distance were frozen when calculating the interaction energy. Our choice has the following reasoning. At the distance ranges of our interest the interaction between the W-tip and sample is rather strong, meaning that the structures of the W-cluster as well as graphene will be significantly modified compared to the isolated cases. In order to simulate experimental conditions we need to consider the interaction between already modified tip and sample. The latter is possible with the chosen approach. If the structures of isolated tip and sample are fully relaxed (i.\,e. the first approach is employed), then the interaction between unmodified sub-systems is investigated. This way may be suitable when the adsorption of free cluster on top of some substrate is investigated.

For each point of the interaction-energy curve (Fig.\,1a, top panel), a vertical distance between the centres of the W cluster and the graphene/Ru(0001) slab is fixed and set to $z = 925\dots1600$\,pm (in total, 27 distances for the ATOP place and 20 distances for the FCC place were considered). Then the coordinates of atoms of the intermediate region are relaxed.

As far as our structural model is completely symmetric with respect to the middle-layer of the slab and middle-layer of the W-cluster, the resulting interaction energy can be calculated according to the formula
\begin{equation}
\Delta E_{\mathrm{int}}(z)=\frac{E_{\mathrm{tip+sample}}(z)-[E_{\mathrm{sample}}(z)+E_{\mathrm{tip}}(z)]}{2}.
\end{equation}
\\
\noindent
\textbf{Calculation of the resonance frequency of graphene nanobubbles.} In order to estimate the graphene stiffness as well as the mechanical resonance frequency of graphehe nanobubbles at the ATOP positions the following procedure was applied. (i) Long-range tail of the experimentally obtained interaction force curves for all high-symmetry positions was fitted with the power function and then this tail was subsequently subtracted from the corresponding total force yielding the curves for the short-range interaction force. Such curve for ATOP position is shown in Fig.\,S7a. (ii) The first minima of this curve gives the reaction of the elastic graphene nanobubbles on the intending tip and if tip was not moved further towards to the substrate then it would describe the adsorption of the metallic cluster at the ATOP position. Therefore, the linear fit of the repulsive part of this curve, in the first approximation, can give the stiffness of graphene at the ATOP position. Such fit yields the stiffness obtained from the experimental data: $k_{exp}=6.88$\,N/m. (iii) The obtained stiffness can be used for the calculation of the Young's modulus of graphene and the respective resonance frequency, using the formula for the resonance frequency of the clamped disc: $k=16\pi E h^3/3R^2$ and $f=0.95\sqrt{E/\rho}$, where $E$ is the Young's modulus, $h=0.335$\,nm is the thickness of graphene, $R=1$\,nm is the radius of the graphene disc clamped at the edges, $\rho =2\times10^3$\,kg/m$^3$ is the density of graphene. These formulas give $E_{exp}=10.9$\,GPa and $f_{exp}=0.74$\,THz. (iv) Similar analysis for the theoretical data (Fig.\,S7b) yield $E_{th}=16.8$\,GPa and $f_{exp}=0.92$\,THz. The obtained, from experiment and theory, values for the Young?s modulus are lower compared to the value ($E=1$\,TPa) obtained in the experiments on the indentation of the micrometer sized graphene film by the AFM tip [C. Lee \textit{et al.}, Science \textbf{321}, 385 (2008)]. These results can open a discussion about validity of mechanics of the continuous media, but can also give a first estimation of these graphene nanoresonators. 

\clearpage
\begin{figure}[h]
  \includegraphics[width=\textwidth]{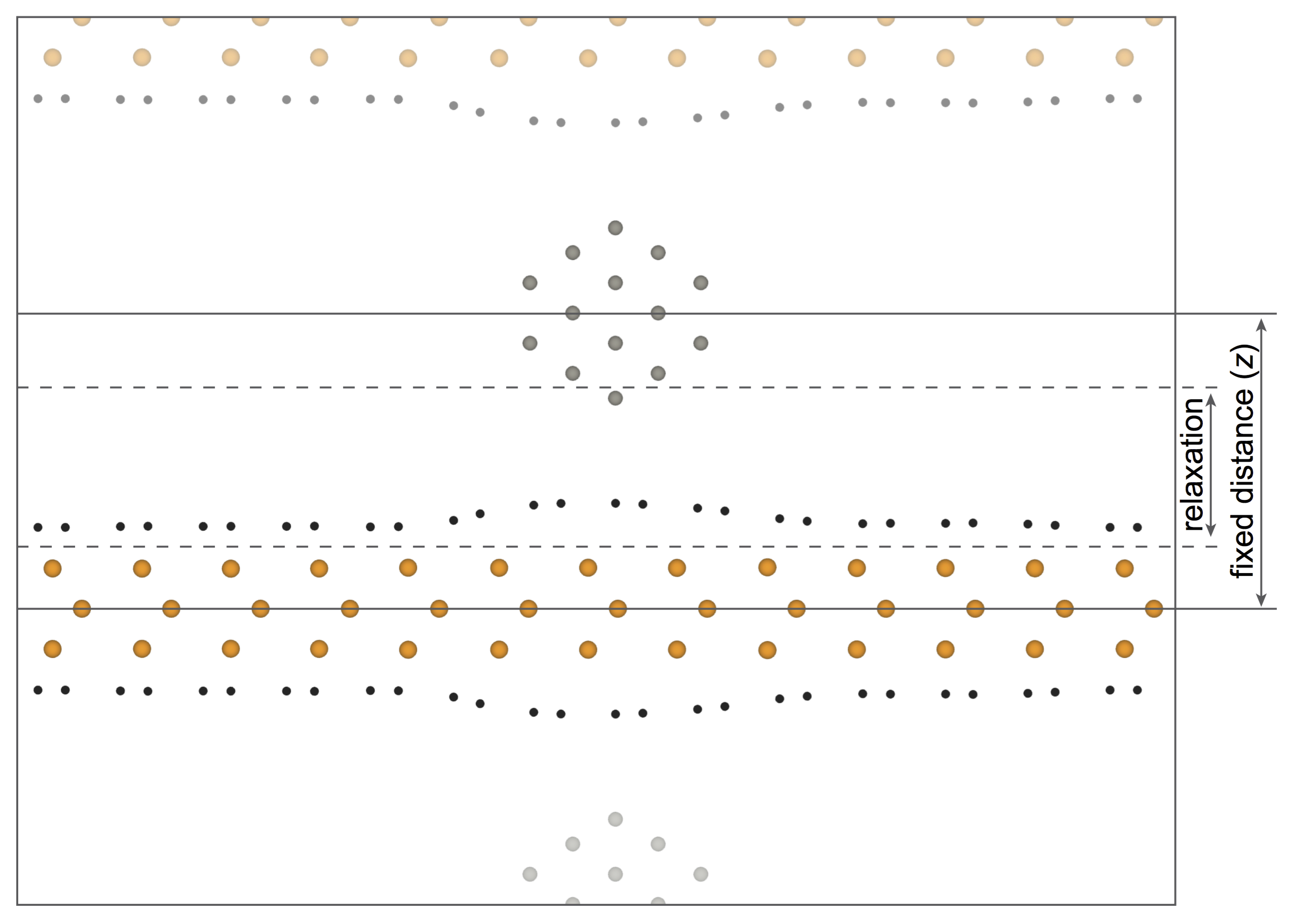}\\
\label{FigS1_SUPPL_slabmodel}
\end{figure}
\noindent\textbf{Fig.\,S1.} Schematic representation of the geometry of the W-tip/graphene/Ru(0001) model used in the simulations of the interaction energy curves.

\clearpage
\begin{figure}[h]
  \includegraphics[width=\textwidth]{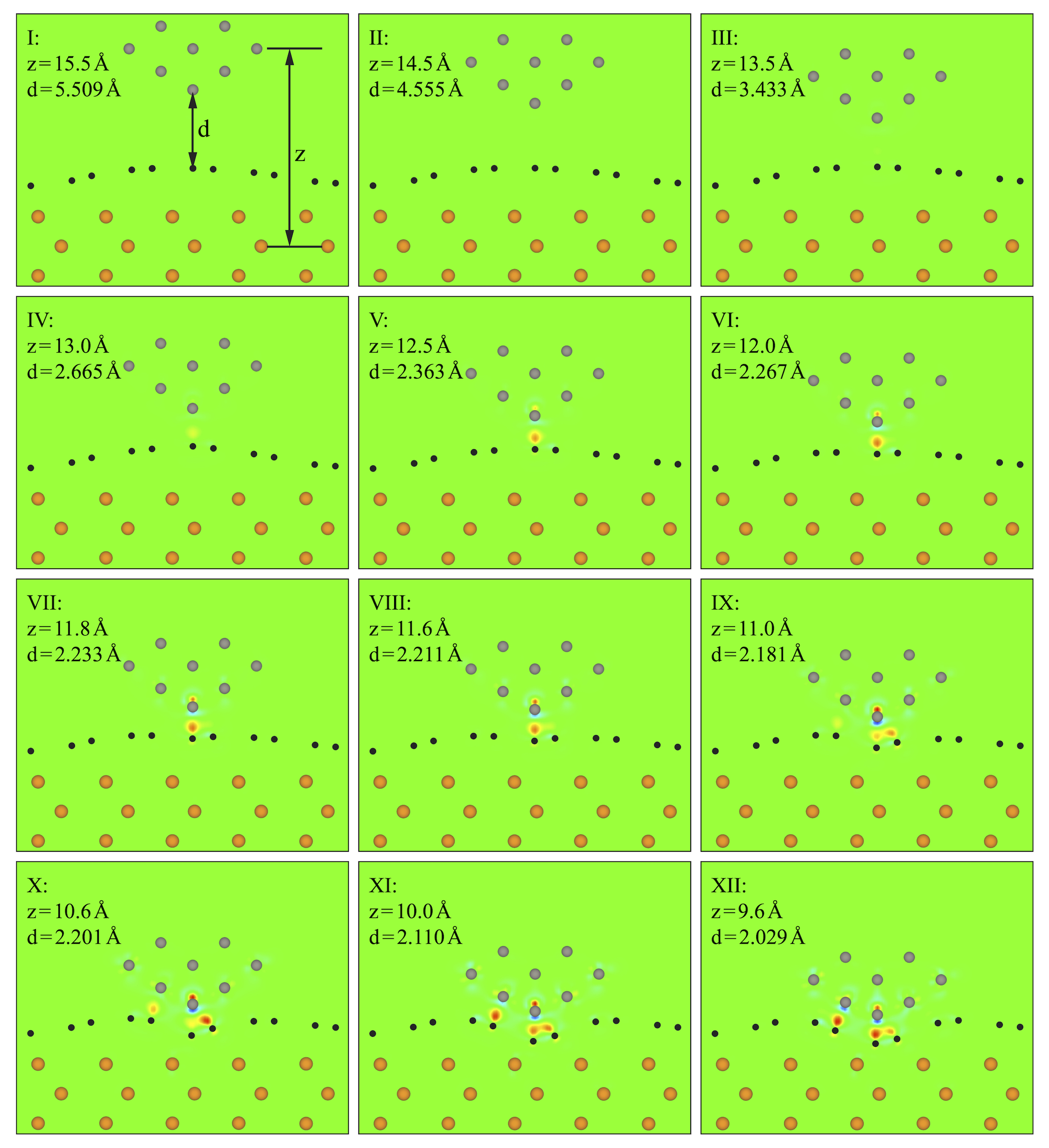}\\
\label{FigS2_SUPPL_snapshots}
\end{figure}
\noindent\textbf{Fig.\,S2.} Extended series of snapshots (similar to the one from Fig.1b). Distance between middle of the model tip and middle of the graphene/Ru(0001) slab ($z$) as well as the distance between W-atom at the apex of the tip and the C-atom directly underneath ($d$) are marked in every figure.

\clearpage
\begin{figure}[h]
  \includegraphics[width=\textwidth]{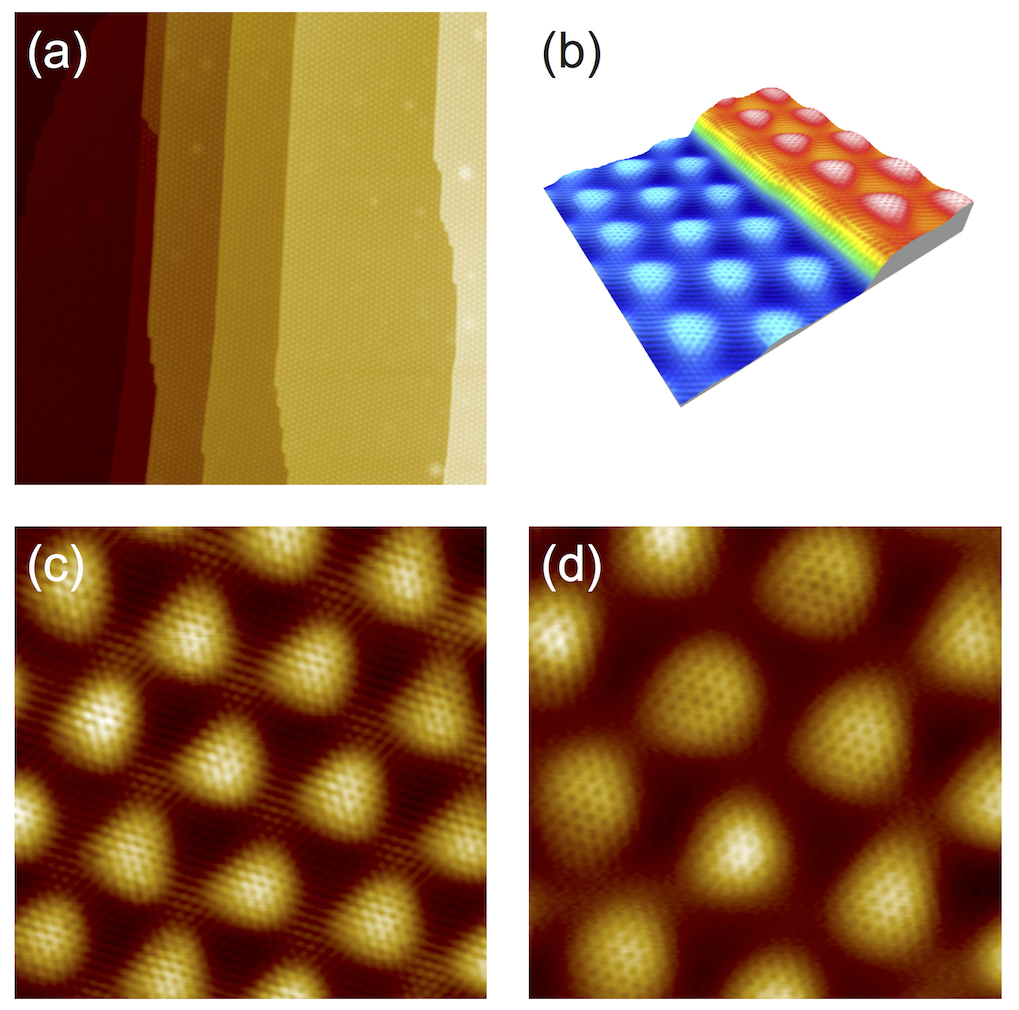}\\
\label{FigS3_SUPPL_STM_AFM}
\end{figure}
\noindent\textbf{Fig.\,S3.} Large and small scale STM (a-c) and nc-AFM (d) images of graphene/Ru(0001). Imaging parameters: (a) $235\times235\,\mathrm{nm}^2$, $U_T = +0.3$\,V, $I_T = 1$\,nA, (b) $13.2\times13.2\,\mathrm{nm}^2$, $U_T = +0.15$\,V, $I_T = 5$\,nA, (c) $10.6\times10.6\,\mathrm{nm}^2$, $U_T = +0.05$\,V, $I_T = 2$\,nA, (d) $8.5\times8.5\,\mathrm{nm}^2$, $U_T = +0.05$\,V, $\Delta f = -6.55$\,Hz, $A = 300$\,pm.

\clearpage
\begin{figure}[h]
  \includegraphics[width=0.5\textwidth]{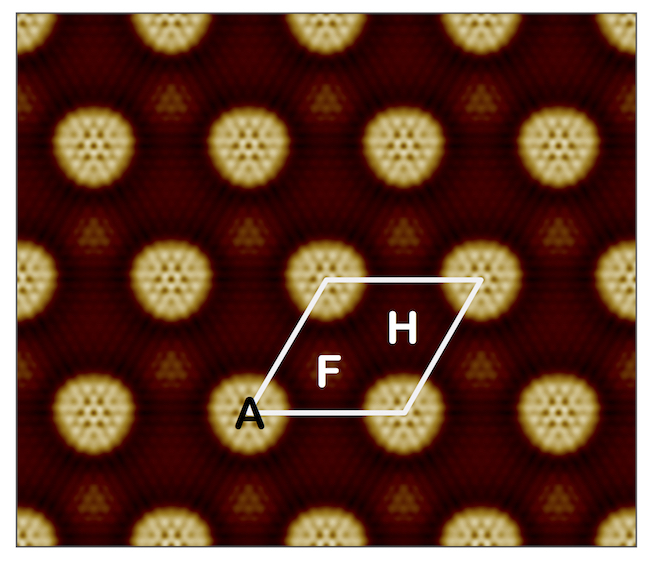}\\
\label{FigS4_SUPPL_STMtheory}
\end{figure}
\noindent\textbf{Fig.\,S4.} Calculated constant current STM image of graphene/Ru(0001) obtained at $+0.3$\,V of the bias voltage. The high-symmetry places are marked by the respective capital letter.

\clearpage
\begin{figure}[h]
  \includegraphics[width=\textwidth]{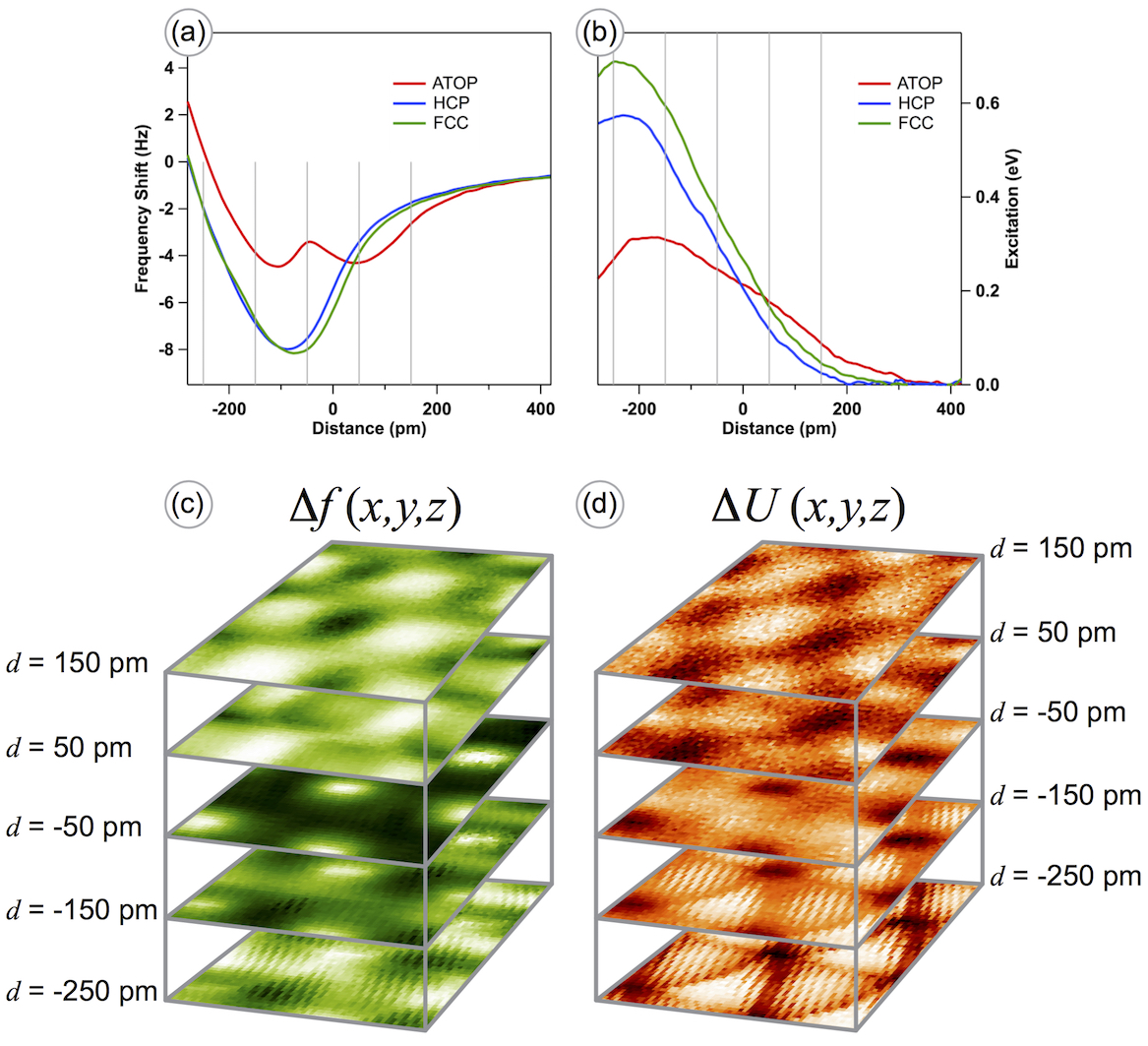}\\
\label{FigS5_SUPPL_3D_AFM}
\end{figure}
\noindent\textbf{Fig.\,S5.} Data extracted from the 3D data sets obtained in the $z$-spectroscopy measurements on the $xy$-grid ($96\,px\times96\,px$), $A = 200$\,pm. (a) and (b) show single  and  curves, respectively, extracted for the middle positions of the respective high-symmetry places of the graphene/Ru(0001) structure. (c) and (d) present series of the constant height $xy$-cuts extracted from the 3D sets of data at the particular distances between tip and graphene (values of $d$ correspond to the respective vertical lines in (a) and (b)). See Movie\,S2 and Movie\,S3 for the complete 3D data sets.

\clearpage
\begin{figure}[h]
  \includegraphics[width=\textwidth]{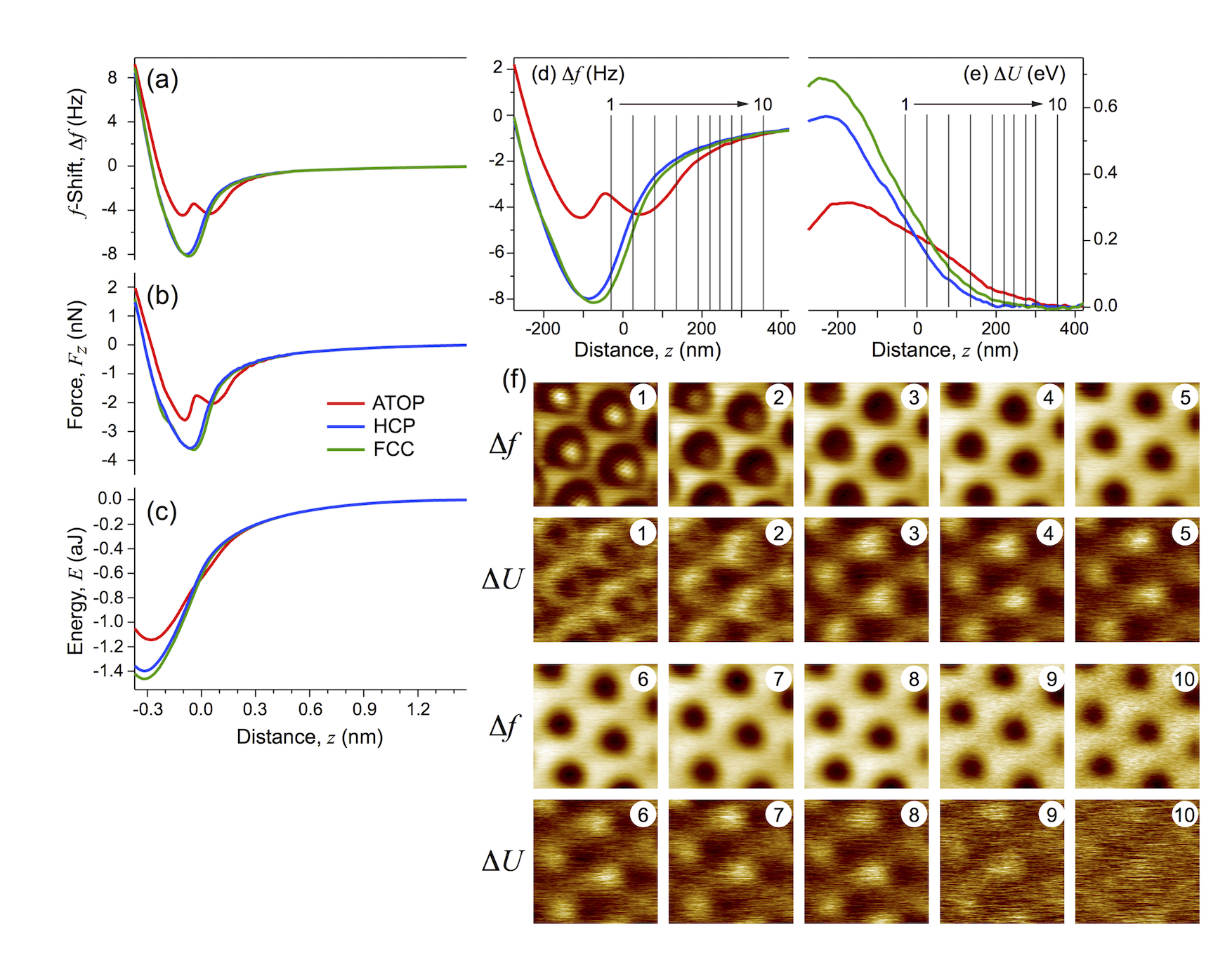}\\
\label{FigS6_SUPPL_df_F_U_exc_CH}
\end{figure}
\noindent\textbf{Fig.\,S6.} (Extended version of Fig.\,2) Force-distance spectroscopy and constant-height SPM of graphene/Ru(0001). (a-c) Frequency shift ($\Delta f$), acting force on the tip ($F_z$), and the tip-graphene interaction energy ($E$) as a function of the tip-sample distance measured at different places of graphene/Ru(0001. (d,e) Frequency shift ($\Delta f$) and dissipation ($\Delta U$) signals zoomed for the tip-sample distances corresponding to the short-range interactions. (f) Series of constant-height SPM images ($\Delta f$ and $\Delta U$ signals) measured at the $z$-positions marked in (d) and (e). Minimum (dark) and maximum (bright) limits for the respective measurement channel can be extracted from the panels (d) and (e). Imaging parameters: $6.4\times6.4\,\mathrm{nm}^2$, $U_T = +0.05$\,V, $A = 200$\,pm.

\clearpage
\begin{figure}[h]
  \includegraphics[width=\textwidth]{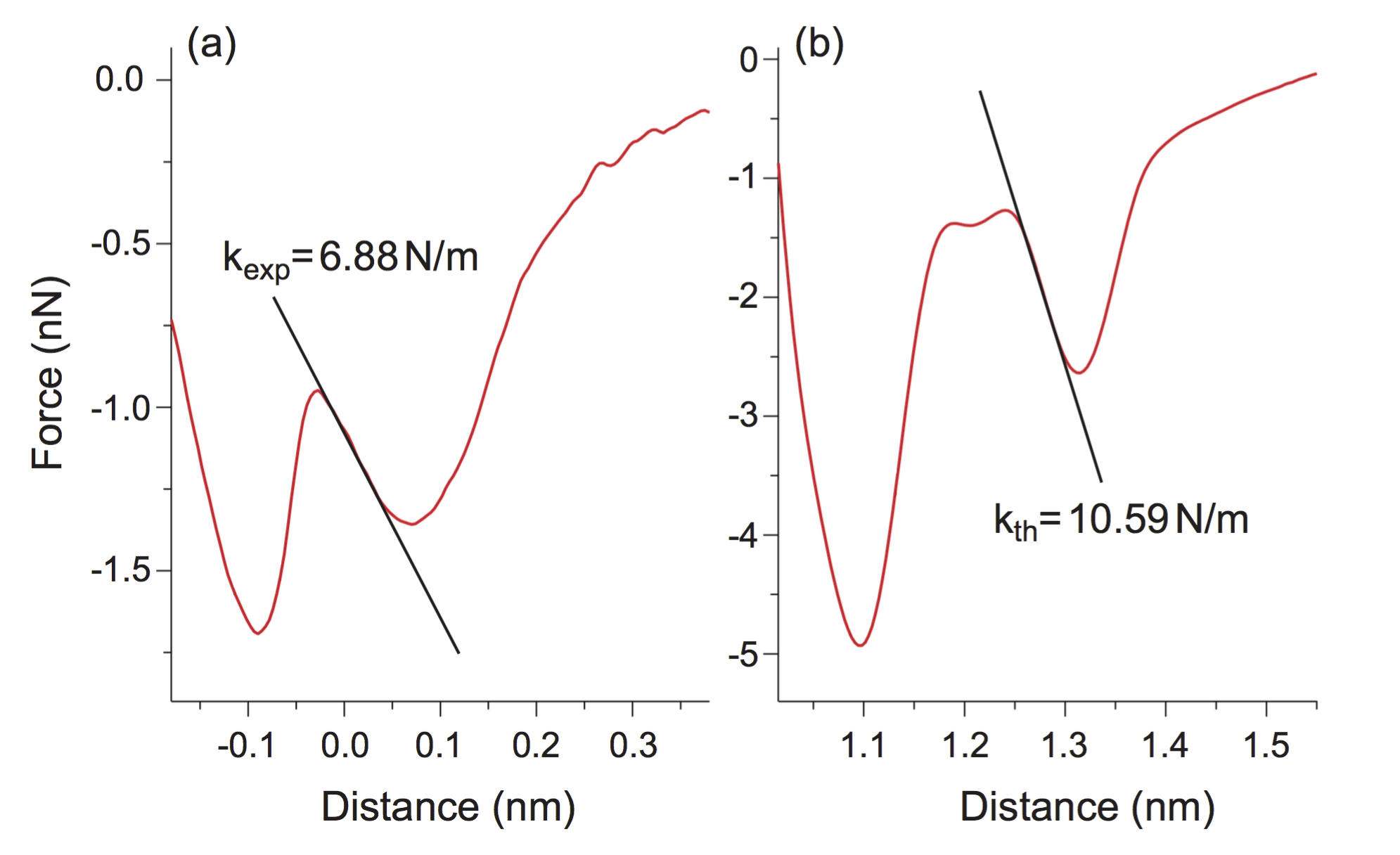}\\
\label{FigS7_SUPPL_f_res}
\end{figure}
\noindent\textbf{Fig.\,S7.} (a) Short-range part of the experimentally obtained interaction force between tip and graphene at the ATOP position after subtraction of the long-range tail. (b) Theoretically calculated interaction force between tip and graphene at the ATOP position. Every panel contains a linear fit of the repulsive part of the first minima and the respective calculated stiffness of graphene.

\end{document}